\documentclass[prl,twocolumn,showpacs,floatfix]{revtex4}

\usepackage{amsmath,amssymb,amsthm}
\usepackage{times}
\usepackage{graphicx}
\usepackage{bm}

\newcommand{\leftindex}{\alpha}

\newcommand{\WFini}{\vert \psi_{ini}\rangle}
\newcommand{\WFtime}{\vert \psi(\tau)\rangle}

\newcommand{\EigenvO}{\vert\Psi_\leftindex^0\rangle}
\newcommand{\Eigenvleft}{\vert\Psi_\leftindex\rangle}

\begin{document}

\title{Breakdown of thermalization in finite one-dimensional systems}

\author{Marcos Rigol} 
\affiliation{Department of Physics, Georgetown University, Washington, DC 20057, USA} 

\pacs{03.75.Kk, 03.75.Hh, 05.30.Jp, 02.30.Ik}

\begin{abstract}
We use quantum quenches to study the dynamics and thermalization of hardcore 
bosons in finite one-dimensional lattices. We perform exact diagonalizations and find 
that, far away from integrability, few-body observables thermalize. 
We then study the breakdown of thermalization as one approaches an integrable point. 
This is found to be a smooth process in which the predictions of standard statistical 
mechanics continuously worsen as the system moves toward integrability. We establish 
a direct connection between the presence or absence of thermalization and the validity 
or failure of the eigenstate thermalization hypothesis, respectively.
\end{abstract}

\maketitle


Little more than fifty years ago, Fermi, Pasta, and Ulam (FPU) \cite{fermi55} 
set up a numerical experiment to prove the ergodic hypothesis for a 
one-dimensional (1D) lattice of harmonic oscillators once nonlinear couplings were 
added. Much to their surprise, the system exhibited long-time periodic dynamics with 
no signals of ergodicity. This behavior could not be explained in terms of Poincar\'e 
recurrences and motivated intense research \cite{focussissue}, which ultimately 
gave rise to the modern chaos theory. It led to the 
discovery of solitons (stable solitary waves) in nonlinear systems and to the 
understanding of thermalization in terms of dynamical chaos \cite{focussissue}. 
In the latter scenario, there is a threshold below which the interactions breaking 
integrability are ineffective in producing chaotic behavior and the system cannot 
be described by standard statistical mechanics \cite{chirikov60}. The FPU numerical 
calculations happened to be below that threshold \cite{izrailev70}.

More recently, experiments with ultracold gases in 
1D geometries have challenged our understanding of the quantum 
domain \cite{kinoshita06}. After bringing a nearly isolated system out of 
equilibrium, no signals of relaxation to the expected thermal equilibrium 
distribution were observed. Some insight can be gained in the framework of 
integrable quantum systems \cite{rigol07STATa}, but then it remains 
the question of why thermalization did not occur even when the system was 
supposed to be away from integrability. In the latter regime, thermalization
is expected to occur \cite{deutsch91,rigol08STATc}. This new experimental
result \cite{kinoshita06} has opened many questions such as: Will thermalization 
occur if one waits longer? Is there a threshold after which thermalization 
will occur? In this work we address some of these questions using numerical 
experiments.

In the limit in which the quantum system is integrable, it has been shown 
numerically \cite{rigol07STATa} that observables such as the ones measured 
experimentally relax to an equilibrium distribution different from the thermal 
one. That a novel distribution is generated is the result of the conserved 
quantities that render the system integrable, and can be characterized by a 
generalization of the Gibbs ensemble (GGE) \cite{rigol07STATa}. Several works 
since then have addressed the relevance and limitations of the GGE to various 
integrable systems and classes of observables \cite{integrable}. 
Much less is known away from integrability where fewer analytical tools are 
available and numerical computations become more demanding. Early works in 
1D have provided mixed results; thermalization was observed in some regimes 
and not in others \cite{kollath07}. In two dimensions, thermalization was 
unambiguously shown to occur \cite{rigol08STATc} and could be understood 
on the basis of the eigenstate thermalization hypothesis \cite{deutsch91}. 
Recent works have also pointed out a possible intermediate quasi-steady 
regime that could occur before thermalization in a class of fermionic systems 
\cite{moeckel08}. Here we study how breaking integrability affects the 
thermalization of correlated bosons in a 1D lattice after a quantum quench.


We consider impenetrable bosons in a periodic 1D lattice with 
nearest-neighbor hopping $t$ and repulsive interaction $V$, and 
next-nearest-neighbor hopping $t'$ and repulsive interaction $V'$. 
The Hamiltonian reads \cite{info}
{\setlength\arraycolsep{0.5pt}
\begin{eqnarray}
\hat{H}&=&\sum_{i=1}^L \left\lbrace -t\left( \hat{b}^\dagger_i \hat{b}_{i+1} + 
\textrm{H.c.} \right) 
+V\left( \hat{n}_i-\dfrac{1}{2}\right)\left( \hat{n}_{i+1}-\dfrac{1}{2}\right) \right.\nonumber \\
&-&\left.t'\left( \hat{b}^\dagger_i \hat{b}_{i+2} + \textrm{H.c.} \right)  
+V'\left( \hat{n}_i-\dfrac{1}{2}\right)\left( \hat{n}_{i+2}-\dfrac{1}{2}\right)
\right\rbrace. 
\end{eqnarray}}

When $t'=V'=0$ this model is integrable. In order to understand how the proximity to the integrable point 
affects equilibration, we prepare an initial state that is an eigenstate of a system with 
$t=t_{ini}$, $V=V_{ini}$, $t'$, $V'$ and then quench the nearest-neighbor parameters to 
$t=t_{fin}$, $V=V_{fin}$ without changing $t'$, $V'$, i.e., we only change 
$t_{ini},V_{ini}\rightarrow t_{fin},V_{fin}$. The same quench is then repeated for 
different values of $t',V'$ as one approaches $t'=V'=0$. We have performed the exact time 
evolution of up to eight impenetrable bosons in lattices with up to 24 sites. Taking 
advantage of translational invariance, this required the full diagonalization of blocks in the
Hamiltonian that contained up to 30,667 states.

Does integrability, or its absence, affect the relaxation dynamics of experimentally 
relevant observables? To answer that question, we examine two of those observables: 
the momentum distribution function $n(k)$ and the structure factor for the 
density-density correlations $N(k)$ \cite{info}. Since the 
initial state wavefunction can be expanded in the eigenstate basis of the final 
Hamiltonian $\widehat{H}$ as $\WFini=\sum_\alpha C_{\alpha} \Eigenvleft$, one finds 
that, if the spectrum is nondegenerate and incommensurate, the infinite time 
average of an observable $\hat{O}$ can be written as
\begin{eqnarray}
\overline{\langle \widehat{O} \rangle} \equiv O_{diag}=\sum_{\alpha} |C_{\alpha}|^{2}
O_{\alpha\alpha},
\label{diagonal}
\end{eqnarray}
where $O_{\alpha\alpha}$ are the matrix elements of $\hat{O}$ in the basis of the final 
Hamiltonian. This exact result can be thought as the prediction of a ``diagonal ensemble'', 
where $|C_{\alpha}|^{2}$ is the weight of each state of this ensemble 
\cite{rigol08STATc}. We then study the normalized area between our observables, during the 
time evolution, and their infinite time average, i.e., at each time we compute
\begin{equation}
 \delta n_k=\dfrac{\sum_k|n(k)-n_{diag}(k)|}{\sum_k n_{diag}(k)}
\label{error}
\end{equation}
and similarly for $\delta N_k$. If, up to small fluctuations, $n(k)$ and $N(k)$ relax to 
a constant distribution, it must be the one predicted by the infinite time average
in Eq.\ (\ref{diagonal}), i.e., $\delta n(k),\,\delta N(k)\rightarrow 0$.

\begin{figure}[!h]
\begin{center}
\includegraphics[width=0.49\textwidth,angle=0]{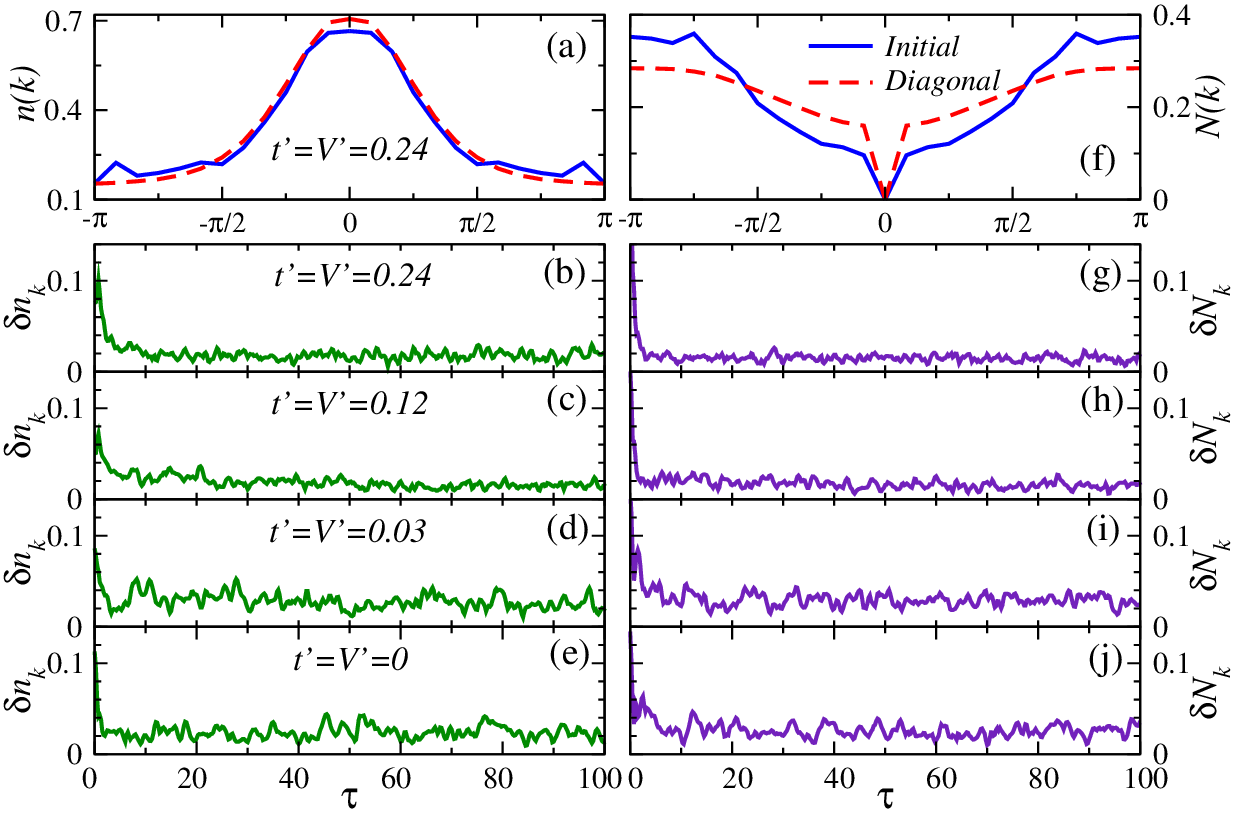}
\end{center}
\vspace{-0.6cm}
\caption{\label{TimeEvolution_L24T2.0}
Quantum quench from $t_{ini}=0.5$, $V_{ini}=2.0$ to $t_{fin}=1.0$, $V_{fin}=1.0$, 
with $t'_{ini}=t'_{fin}=t'$ and $V'_{ini}=V'_{fin}=V'$ in a system with eight
hardcore bosons ($N_b=8$) in 24 lattice sites ($L=24$). The initial state was chosen 
within the eigenstates of the initial Hamiltonian with total momentum $k=0$ in such 
a way that after the quench the system has an effective temperature $T=2.0$ in all 
cases. Given the energy of the initial state in the final Hamiltonian 
$E=\langle\psi_{ini}\vert \widehat{H}_{fin}\vert \psi_{ini}\rangle$ the effective temperature 
is computed as described in Ref.\ \cite{temperature}. 
(a) Initial and diagonal ensemble results for $n(k)$ when $t'=V'=0.24$. 
(b)--(e) Time evolution of $\delta n_k$ for $t'=V'=0.24$, 0.12, 0.03, and 0.0 respectively. 
(f) Initial and diagonal ensemble results for $N(k)$ when $t'=V'=0.24$. 
(g)--(j) Time evolution of $\delta N_k$ for $t'=V'=0.24$, 0.12, 0.03, and 0.0 respectively.}
\end{figure}

In Fig.\ \ref{TimeEvolution_L24T2.0}, we show results for $\delta n_k$ and $\delta N_k$ 
as a function of time $\tau$ for four different quenches as one approaches the 
integrable point. In Figs.\ \ref{TimeEvolution_L24T2.0}(a) and 
\ref{TimeEvolution_L24T2.0}(f), we compare the initial $n(k)$ 
and $N(k)$ with the predictions of Eq.\ (\ref{diagonal}) away from 
integrability. Interestingly, the time evolution of $\delta n_k$ and $\delta N_k$ in 
Figs.\ \ref{TimeEvolution_L24T2.0}(b)-\ref{TimeEvolution_L24T2.0}(e) and
Figs.\ \ref{TimeEvolution_L24T2.0}(g)-\ref{TimeEvolution_L24T2.0}(j), respectively, 
can be seen to be very similar for all values of $t'$ and $V'$. Hence, the dynamics 
are barely affected by the closeness to the integrable point \cite{info}. 
In all cases, we find that there is a fast relaxation of $n(k)$ and $N(k)$ 
towards the diagonal ensemble prediction (in a time scale $\tau\sim 1/t_{fin}$ \cite{info}) 
and that later they fluctuate within one to two percent of that prediction. The latter 
differences decrease as one increases the system size and the energy, or the effective 
temperature \cite{temperature} of the system \cite{info}. From these results, 
we infer that, for very large systems sizes and finite effective temperatures, 
$n(k)$ and $N(k)$ should in general relax to exactly the predictions of 
Eq.\ (\ref{diagonal}) even if the system is very close or at integrability. 

\begin{figure}[!b]
\begin{center}
\includegraphics[width=0.41\textwidth,angle=0]{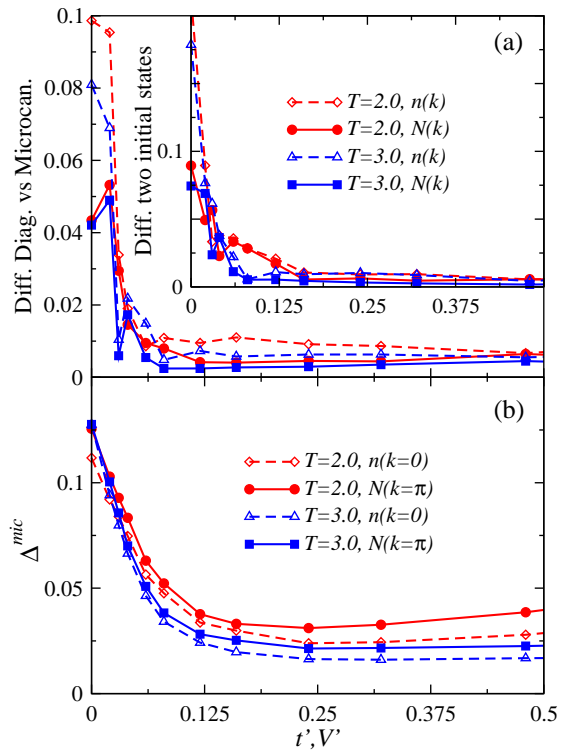}
\end{center}
\vspace{-0.6cm}
\caption{\label{Thermodynamics}
Comparison between the predictions of different statistical ensembles 
for $n(k)$ and $N(k)$ after relaxation in a system with $L=24$ and $N_b=8$. 
Results are shown for $T=2.0$ and $T=3.0$. 
(a) (main panel) Relative difference between the predictions of the 
diagonal ensemble and the microcanonical ensemble.
(a) (inset) Relative difference between the predictions of two 
diagonal ensembles generated by different initial states selected from 
the eigenstates of a Hamiltonian with $t_{ini}=0.5$, $V_{ini}=2.0$ and a 
Hamiltonian with $t_{ini}=2.0$, $V_{ini}=0.5$, and $L=24$, $N_b=8$. The 
final Hamiltonian ($t_{fin}=1.0$, $V_{fin}=1.0$) and the effective 
temperature are identical for both initial states.
Relative differences are computed following Eq.\ (\ref{error}).
(b) Average relative deviation of eigenstate expectation values with 
respect to the microcanonical prediction (see text) as a function of $t',V'$, 
for the same effective temperatures as in (a).}
\end{figure}

Once it is known that the diagonal ensemble provides a very good description of 
observables after their relaxation dynamics, a question that remains to be 
answered is how good are standard statistical ensembles in reproducing the 
$n(k)$ and $N(k)$ distributions predicted by the diagonal ensemble as one 
approaches the integrable point? Based on the experimental results \cite{kinoshita06} 
one would expect conventional statistical mechanics to fail everywhere in 1D or 
at least over a finite window in the vicinity of the integrable point. We find 
the latter scenario to be the one realized in our finite 1D lattices.

In the main panel of Fig.\ \ref{Thermodynamics}(a), we compare the diagonal ensemble 
results with the predictions of the microcanonical ensemble for our two observables of 
interest. We find the differences between them to be below one percent and decreasing 
with system size \cite{info} when the system is far from 
integrability ($t'=V'>0.1$). Thus, one can say that thermalization takes place in this case. 
As one approaches the integrable point, on the other hand, the differences between the 
diagonal and microcanonical ensembles increase, signaling a breakdown of thermalization in 
1D. This breakdown is accompanied by a dependence of the properties of the 
system after relaxation on the initial state, as can be seen in the inset in 
Fig.\ \ref{Thermodynamics}(a). There, we compare the predictions of the diagonal ensemble 
for two different initial states that have the same effective temperature. 
Those predictions are almost independent of the initial 
state for $t'=V'>0.1$, but strongly dependent on it closer to integrability.
In Fig.\ \ref{Thermodynamics}(a), it is remarkable to find that both $n(k)$ and $N(k)$ 
exhibit very similar quantitative behavior away from integrability, which is an 
indication of the generality of our results for few-body observables.

\begin{figure}[!h]
\begin{center}
\includegraphics[width=0.4\textwidth,angle=0]{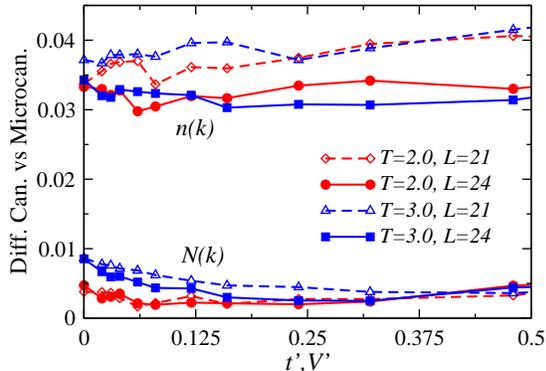}
\end{center}
\vspace{-0.6cm}
\caption{\label{Microvscanonical}
Relative difference between the predictions of the microcanonical 
and canonical ensembles for our two observables of interest in systems 
with $L=21$, $N_b=7$ and $L=24$, $N_b=8$. Relative differences are computed 
following Eq.\ (\ref{error}).}
\end{figure}

A clarification is in order at this point. It is usually assumed that, for extensive 
quantities, the predictions of conventional statistical mechanical ensembles such as the 
microcanonical and canonical ones are identical, provided the systems are large enough. 
We should stress that this is not the case in our small systems, and may not be 
the case for many experiments that are performed in similar setups. In 
Fig.\ \ref{Microvscanonical} one can see that depending on the physical observable 
under consideration the differences between the predictions of those ensembles can 
be quite large [in particular for $n(k)$]. Interestingly, they are almost not affected 
by the proximity to integrability, and, as expected, can be seen to reduce 
with increasing system size.

\begin{figure}[!h]
\begin{center}
\includegraphics[width=0.43\textwidth,angle=0]{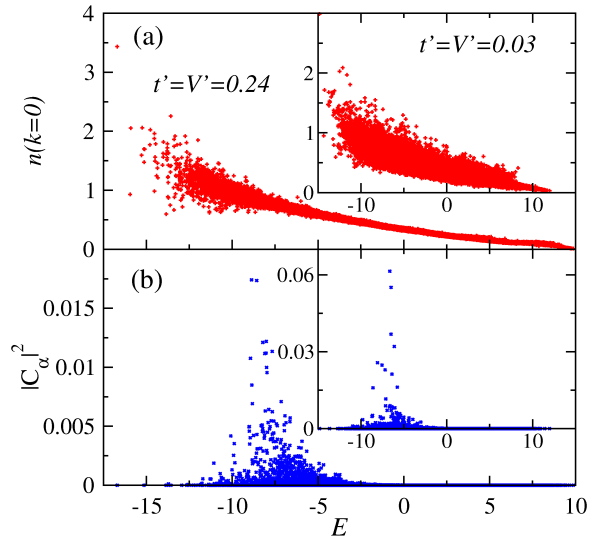}
\end{center}
\vspace{-0.6cm}
\caption{\label{ETH}
(a) $n(k=0)$ as a function of the energy for all the eigenstates of the 
Hamiltonian (including all momentum sectors). 
(main panel) $t=V=1$ and $t'=V'=0.24$. (inset) $t=V=1$ and $t'=V'=0.03$. 
The system has 24 lattice sites and 8 bosons for which the total Hilbert space 
consists of 735,471 states.
(b) Distribution of $|C_{\alpha}|^{2}$ for the quench from 
$t_{ini}=0.5$, $V_{ini}=2.0$ to $t_{fin}=1.0$, $V_{fin}=1.0$ for an 
effective temperature $T=2.0$. (main panel) $t=V=1$ and $t'=V'=0.24$, 
where $E=\langle\psi_{ini}\vert \widehat{H}_{fin}\vert \psi_{ini}\rangle=-6.85$. 
(inset) $t=V=1$ and $t'=V'=0.03$, where 
$E=-5.95$.}
\end{figure}

One may wonder why conventional statistical mechanical ensembles can 
predict the exact values of few-body observables after relaxation at all, 
since these seem to be a priori fully dependent on the initial conditions through 
the values of $|C_{\alpha}|^{2}$ [see Eq.\ (\ref{diagonal})]. In the main panel 
in Fig.\ \ref{ETH}(b), we show the distribution of $|C_{\alpha}|^{2}$ for one of 
the quenches that exhibited thermalization in Fig.\ \ref{Thermodynamics}(a). 
Figure \ref{ETH}(b) shows that the $|C_{\alpha}|^{2}$ distribution is clearly 
different from the microcanonical one, in which all states within an energy window 
are taken with the same weight. One can conclude then that something else is 
at play here.

A resolution to this puzzle was advanced by Deutsch and Srednicki 
in terms of the eigenstate thermalization hypothesis (ETH) \cite{deutsch91}.
ETH states that the fluctuation of eigenstate expectation values of generic few-body 
observables [$O_{\alpha\alpha}$ in Eq.\ (\ref{diagonal})] is small between
eigenstates that are close in energy, which means that the microcanonical average is identical 
to the prediction of each eigenstate, or what is the same that the eigenstates already 
exhibit a thermal value of the observable. If this holds, thermalization in an isolated quantum 
system will follow for any distribution of $|C_{\alpha}|^{2}$, as long as it is narrow 
enough in energy. This scenario was shown to be valid in isolated two-dimensional 
systems in Ref.\ \cite{rigol08STATc}.

In 1D systems, we find the onset of thermalization to be directly related 
to the validity of the eigenstate thermalization hypothesis. In the main panel in 
Fig.\ \ref{ETH}(a), we depict $n(k=0)$ [similar results were obtained for $n(k\neq0)$ 
and $N(k)$] in each eigenstate of the Hamiltonian as a function of the energy of the 
eigenstate, when the system is far from integrability. 
Figure \ref{ETH}(a) shows that after a region of low energies where the eigenstate 
expectation values exhibit large fluctuations follows another region where fluctuations 
are small and ETH holds. The inset in Fig. \ref{ETH}(a) shows that for a system close 
to the integrable point, in which thermalization is absent [Fig.\ \ref{Thermodynamics}(a)], 
the eigenstate to eigenstate fluctuations of $n(k=0)$ are very large over the entire 
spectrum and ETH does not hold.

\begin{figure}[!h]
\begin{center}
\includegraphics[width=0.49\textwidth,angle=0]{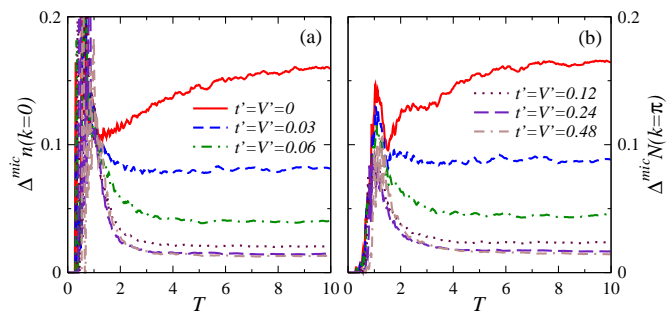}
\end{center}
\vspace{-0.6cm}
\caption{\label{DeviationfromETH}
Average relative deviation of eigenstate expectation values with 
respect to the microcanonical prediction as a function of the 
effective temperature of the microcanonical ensemble. Results are presented 
for: (a) $n(k=0)$ and (b) $N(k=\pi)$. These systems 
have 24 lattice sites, 8 bosons, and $t=V=1.0$. An effective 
temperature $T=2$ in these plots corresponds to an energy 
$E_{mic}=-5.95$ for $t'=V'=0.03$ and 
$E_{mic}=-6.85$ for $t'=V'=0.24$ and,
$T=10$ corresponds to an energy 
$E_{mic}=-0.81$ for $t'=V'=0.03$ and 
$E_{mic}=-0.82$ for $t'=V'=0.24$
(see the corresponding parts of the spectrum in Fig.\ \ref{ETH}).}
\end{figure}

In order to be more quantitative, and to understand how ETH breaks down as one approaches 
the integrable point, we have computed the average relative deviation of the eigenstate 
expectation values with respect to the microcanonical prediction ($\Delta^{mic}$) 
for $n(k=0)$ and for $N(k=\pi)$ \cite{info}. They are shown in 
Fig.\ \ref{DeviationfromETH}(a) and \ref{DeviationfromETH}(b), respectively, as a function
of the effective temperature $T$ corresponding to an energy $E$ of the microcanonical 
ensemble \cite{temperature}. Temperature in this plot is only used as an auxiliary tool 
for assessing how far from the ground state these systems are. 

In Fig.\ \ref{DeviationfromETH} one can see that 
$\Delta^{mic}n(k=0)$ and $\Delta^{mic}N(k=\pi)$ exhibit very similar behavior. 
Below $T\simeq 1.5$ fluctuations are large and due to the relatively small number of 
states in some energy windows the use of the microcanonical ensemble may not be 
well justified. For $T\gtrsim 1.5$, the fluctuations saturate with increasing
temperature, and $\Delta^{mic}n(k=0)$ and $\Delta^{mic}N(k=\pi)$ continuously 
increase, at any given temperature, as one approaches the integrable point. 
The latter is better seen in Fig.\ \ref{Thermodynamics}(b) for the two specific 
effective temperatures studied in Fig.\ \ref{Thermodynamics}(a). A comparison 
between Fig.\ \ref{Thermodynamics}(a) and Fig.\ \ref{Thermodynamics}(b), allows us 
to stablish a direct connection between the breakdown of thermalization in 1D 
and the increase of the eigenstate to eigenstate fluctuations in the Hamiltonian,
or what is the same, the failure of ETH. We have also studied how these results are modified 
when changing the system size. We find that for our lattice sizes $\Delta^{mic}$ 
still reduces as one increases the system size \cite{info}. The relative reduction 
is more pronounced away from integrability.

Overall, our study shows that in finite 1D systems there is a 
regime close to integrability where thermalization fails to occur, and that this 
failure originates in the breakdown of ETH. We also find that ETH holds further 
departing from integrability explaining why thermalization can occur in  
1D. Since both experiments \cite{kinoshita06} and computations have been 
performed with relatively small systems, an important question that remains 
open is whether for sufficiently large system sizes thermalization will occur 
arbitrarily close to the integrable point or whether there will be a finite critical 
value starting from which integrability breaking terms will produce thermalization. 
For our finite systems all we see is a smooth breakdown of thermalization 
as one approaches integrability \cite{smooth}. The answer to this question may be well 
suited for experimental analysis as it may be easier to study the scaling 
with system size in experiments than within our numerical computations, 
which scale exponentially with the system size. The addition of a lattice along 
the 1D tubes in the experiments \cite{paredes04} may help in 
addressing these issues.

\begin{acknowledgments}
This work was supported by startup funds from Georgetown University and by 
the US Office of Naval Research Award No.\ N000140910966. We are grateful to 
Maxim Olshanii for discussions and useful comments on the manuscript.
\end{acknowledgments}

%
%

\onecolumngrid

\vspace*{0.4cm}

\begin{center}

{\large \bf Supplementary material for EPAPS
\\ Breakdown of thermalization in finite one-dimensional systems.}\\

\vspace{0.6cm}

Marcos Rigol\\

{\it Department of Physics, Georgetown University, Washington, DC 20057, USA}
 
\end{center}

\vspace{0.6cm}

\twocolumngrid


\subsection{Hamiltonian and Numerical Calculations}

In a system of units where $\hbar=1$, the Hamiltonian for our model of impenetrable bosons 
(hardcore bosons) in a one-dimensional lattice with periodic boundary conditions reads
{\setlength\arraycolsep{0.5pt}
\begin{eqnarray}
\hat{H}&=&\sum_{i=1}^L \left\lbrace -t\left( \hat{b}^\dagger_i \hat{b}_{i+1} + \textrm{H.c.} \right) 
+V\left( \hat{n}_i-\dfrac{1}{2}\right)\left( \hat{n}_{i+1}-\dfrac{1}{2}\right) \right.\nonumber \\
&-&\left.t'\left( \hat{b}^\dagger_i \hat{b}_{i+2} + \textrm{H.c.} \right)  
+V'\left( \hat{n}_i-\dfrac{1}{2}\right)\left( \hat{n}_{i+2}-\dfrac{1}{2}\right)\right\rbrace. 
\label{hamiltonian}
\end{eqnarray}
}where the hardcore boson creation and annihilation operators at site 
$i$ are denoted by $\hat{b}^{\dagger}_{i}$ and $\hat{b}^{}_{i}$, respectively, 
and the local density operator by $\hat{n}_i=\hat{b}^{\dagger}_{i}\hat{b}^{}_{i}$.
In different sites the creation and annihilation operators
commute as usual for bosons:
\begin{equation}
[\hat{b}^{}_{i},\hat{b}^{\dagger}_{j}]=
[\hat{b}^{}_{i},\hat{b}^{}_{j}]=
[\hat{b}^{\dagger}_{i},\hat{b}^{\dagger}_{j}]=0,\quad \mathrm{for} \quad i\neq j.
\end{equation}
However, on the same site the hardcore bosons operators satisfy anticommutation 
relations typical of fermions:
\begin{equation}  
\left\lbrace  \hat{b}^{}_{i},\hat{b}^{\dagger}_{i}\right\rbrace =1, 
\qquad
\hat{b}^{\dagger 2}_{i}= \hat{b}^2_{i}=0.
\label{ConstHCB} 
\end{equation}
These constraints avoid double or higher occupancy of the lattice sites. 
In Eq.\ (\ref{hamiltonian}), the nearest and next-nearest-neighbor hopping parameters 
are denoted by $t$ and $t'$, the nearest and next-nearest-neighbor interactions 
are denoted by $V$ and $V'$, which in our study are always repulsive, i.e., $V,V'>0$,
and $L$ is the number of lattice sites.

For $t'=V'=0$, the above Hamiltonian is integrable. It also exhibits 
a superfluid to insulator transition at half-filling as $V$ is 
increased [$(V/t)_c=2$]. Finite values of $t'$ and $V'$ break 
integrability and generate a plethora of competing phases. All 
our results in this work are obtained for quenches within the 
superfluid phase.

For our exact study of the nonequilibrium dynamics, we perform 
a full diagonalization of the many-body Hamiltonian (\ref{hamiltonian}) 
taking advantage of the translational symmetry of the lattice. Our initial 
state is always selected from the zero total momentum sector. The 
largest $k=0$ sector in our study corresponds to eight bosons in 24 
lattices sites, where matrices of dimension $D=30,667$ are diagonalized. 
Since our quenches do not break translational symmetry only states with 
$k=0$ are required for the time evolution, and we use all of them
\[
\WFtime=e^{-i\widehat{H}\tau}\WFini= 
\sum_\alpha C_{\alpha}e^{-i E_\alpha \tau}\EigenvO \, ,
\]
where $\WFtime$ is the time-evolving state, $\WFini$ is the initial state, 
$\EigenvO$ are the eigenstates of the Hamiltonian with zero total momentum 
and energy $E_\alpha$, and $C_{\alpha}=\langle \Psi^0_\alpha\WFini$.

The observables of interest in this work are the momentum distribution 
function of the bosons
\[
 \hat{n}(k)=\dfrac{1}{L}\sum_{i,j} e^{-k(i-j)} \hat{b}^{\dagger}_i\hat{b}^{}_j,
\]
which is generally measured in ultracold gases 
experiments, and the density-density correlation structure factor
\[
\hat{N}(k)=\dfrac{1}{L}\sum_{i,j} e^{-k(i-j)} 
\hat{n}_i\hat{n}_j .
\]
Since we always work at a fixed number of particles $N_b$, the expectation value 
of $\hat{N}(k=0)$ is trivially $\langle \hat{N}(k=0)\rangle=N_b^2/L$. 
We subtract this trivial constant in all our plots. 

Studying a one-body and a two-body observable allows us to 
assess the generality of our results for generic few-body observables in
generic one-dimensional systems.

\subsection{Relaxation Dynamics after a Quantum Quench}

\begin{figure}[!h]
\begin{center}
\includegraphics[width=0.7\textwidth,angle=0]{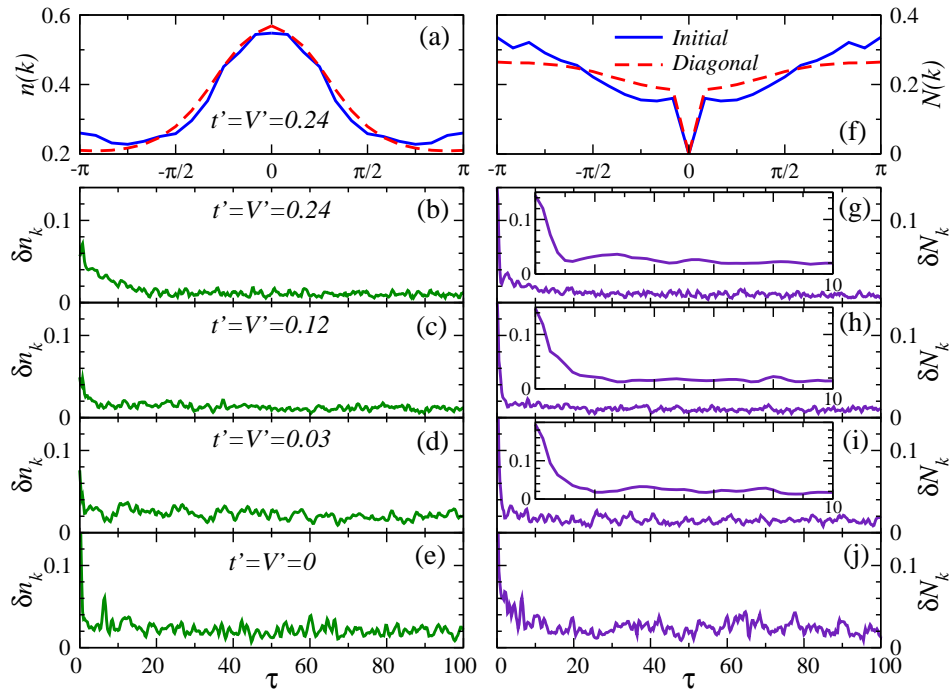}
\end{center}
\vspace{-0.7cm}
\caption{\label{TimeEvolution_L24T3.0} 
Quantum quench from $t_{ini}=0.5$, $V_{ini}=2.0$ to $t_{fin}=1.0$, $V_{fin}=1.0$, 
with $t'_{ini}=t'_{fin}=t'$ and $V'_{ini}=V'_{fin}=V'$ in a system with eight
hardcore bosons ($N_b=8$) in 24 lattice sites ($L=24$). The initial state was chosen 
within the eigenstates of the initial Hamiltonian with total momentum $k=0$ in such 
a way that after the quench the system has an effective temperature $T=3.0$ in all 
cases (see text). (a) Initial and diagonal ensemble results for $n(k)$ and
$t'=V'=0.24$. (b)--(e) Time evolution of $\delta n_k$ for 
$t'=V'=0.24$, 0.12, 0.03, and 0.0 respectively. 
(f) Initial and diagonal ensemble results for $N(k)$ and $t'=V'=0.24$. 
(g)--(j) Time evolution of $\delta N_k$ for $t'=V'=0.24$, 0.12, 0.03, 
and 0.0 respectively. (g)--(i) (insets) Magnification of the short-time 
scales for the time evolution depicted in the main panels.}
\end{figure}

In Fig.\ 1 (main text), we have shown results for the relaxation 
dynamics of a system with 8 bosons ($N_b=8$) in 24 lattice sites ($L=24$). 
The initial state was selected so that the time evolving system has an effective 
temperature $T=2.0$. [Given the energy of the time evolving state in the final 
Hamiltonian, which is conserved $E=\langle\psi_{ini}\vert \widehat{H}_{fin}\vert \psi_{ini}\rangle$, 
the effective temperature $T$ is defined by the expression 
$E=Z^{-1}\textrm{Tr}\left\lbrace {\hat{H}_{fin} \exp(-\hat{H}_{fin}/{k_B T})}\right\rbrace$, 
where $\hat{H}_{fin}$ is the final Hamiltonian, 
$Z=\textrm{Tr}\left\lbrace {\exp(-\hat{H}_{fin}/{k_B T})}\right\rbrace$, 
and $k_B$ the Boltzmann constant. The trace runs over the full spectrum, 
including all momentum sectors.]
In Fig.\ \ref{TimeEvolution_L24T3.0}, we present results for a quench with exactly 
the same Hamiltonian parameters as in Fig.\ 1 (main text) but taking 
a different initial state so that the energy of the system is higher, 
corresponding to an effective temperature $T=3.0$. (Notice that for all our 
results $t_{fin}=1.0$ sets the energy scale.)

Overall, we find that the time scale over which relaxation occurs is of the order 
$\tau \sim 1/t_{fin}$ in all cases studied. The short time scale behavior can be better 
seen in the insets in Fig.\ \ref{TimeEvolution_L24T3.0}(g)-\ref{TimeEvolution_L24T3.0}(i).
In addition, the results in Fig.\ \ref{TimeEvolution_L24T3.0}, for an effective 
$T=3.0$, show that the values of $\delta n_k$ and $\delta N_k$ after relaxation 
are slightly smaller and fluctuate less than the ones obtained for an 
effective temperature $T=2.0$ [Fig.\ 1 (main text)]. This implies 
that increasing the energy of the final state improves the convergence towards 
the infinite time average prediction. 

This difference can be understood in terms of a higher number of eigenstates 
of the Hamiltonian that participate during the nonequilibrium dynamics 
as the energy of the time-evolving state is increased. Since more states 
are involved, dephasing is more effective and time fluctuations after 
relaxation are smaller in these finite systems. The higher participation 
can be understood as the density of states increases with increasing 
energy, so that in general there are more states available close to 
any given energy the higher the latter is.

\onecolumngrid

\twocolumngrid

\begin{figure}[!h]
\begin{center}
\includegraphics[width=0.485\textwidth,angle=0]{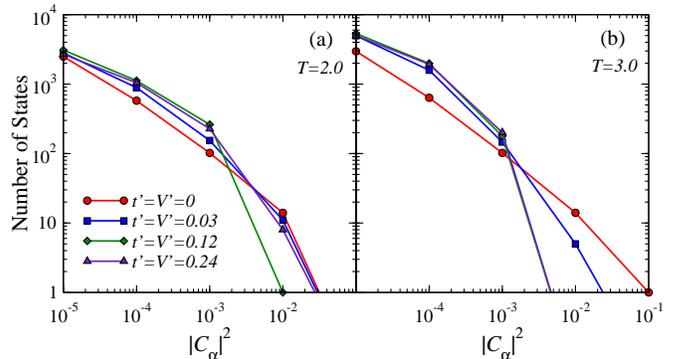}
\end{center}
\vspace{-0.7cm}
\caption{\label{StateCount}
Number of states with $|C_\alpha|^2$ greater than the value presented in the $x$
axis, for an effective temperature (a) $T=2.0$ and (b) $T=3.0$, 
and for exactly same quenches studied in Fig.\ 1 (main text) and 
Fig.\ \ref{TimeEvolution_L24T3.0}, respectively.}
\end{figure}

In Fig.\ \ref{StateCount}, we depict the number of states that have a value of
$|C_\alpha|^2$ greater than $10^{-5},\, 10^{-4},\,\ldots, 0.1$, when the effective
temperature is 2.0 [Fig.\ \ref{StateCount}(a)] and 3.0 [Fig.\ \ref{StateCount}(b)]. 
One can see there that the number of states with $|C_\alpha|^2>10^{-5}$, 
$|C_\alpha|^2>10^{-4}$, and $|C_\alpha|^2>10^{-3}$ is larger for $T=3.0$ than for $T=2.0$
(notice that this is a log-log plot). Since $\sum_\alpha|C_\alpha|^2=1$, this means 
that at lower temperatures there are more states with larger values of 
$|C_\alpha|^2$, but since those are very few in any case, dephasing tends to
be less effective as the temperature decreases.

The same argument that allows us to understand the effect of the effective 
temperature helps in understanding why at integrability, or very close to it, 
$\delta n_k$ and $\delta N_k$ exhibit larger time fluctuations and larger average 
values. The effect is small ($\sim 1\%$) in Fig.\ 1 (main text) and 
Fig.\ \ref{TimeEvolution_L24T3.0}, but can be also understood by analyzing 
Fig.\ \ref{StateCount}. There, one can see that the number of states with the 
largest values of $|C_\alpha|^2$ is always the largest at integrability and 
continuously evolves as one moves away from integrability. 
In Fig.\ \ref{StateCount}(b) such evolution is seen to saturate for $t'=V'>0.12$. The effect
of integrability on the number of states with large $|C_\alpha|^2$ is related to the presence 
of additional conserved quantities, which given the initial conditions, restrict the number 
of states close to the energy of the system that can have a significant overlap with the 
initial state.

We should add that the fact that in all cases time fluctuations are found to be 
relatively small, even though these systems are not only finite but small, can be related 
to the smallness of the off-diagonal elements of the physical observables under 
consideration and has been discussed before \cite{srednicki96,rigol08STATac}.

\begin{figure}[!h]
\begin{center}
\includegraphics[width=0.7\textwidth,angle=0]{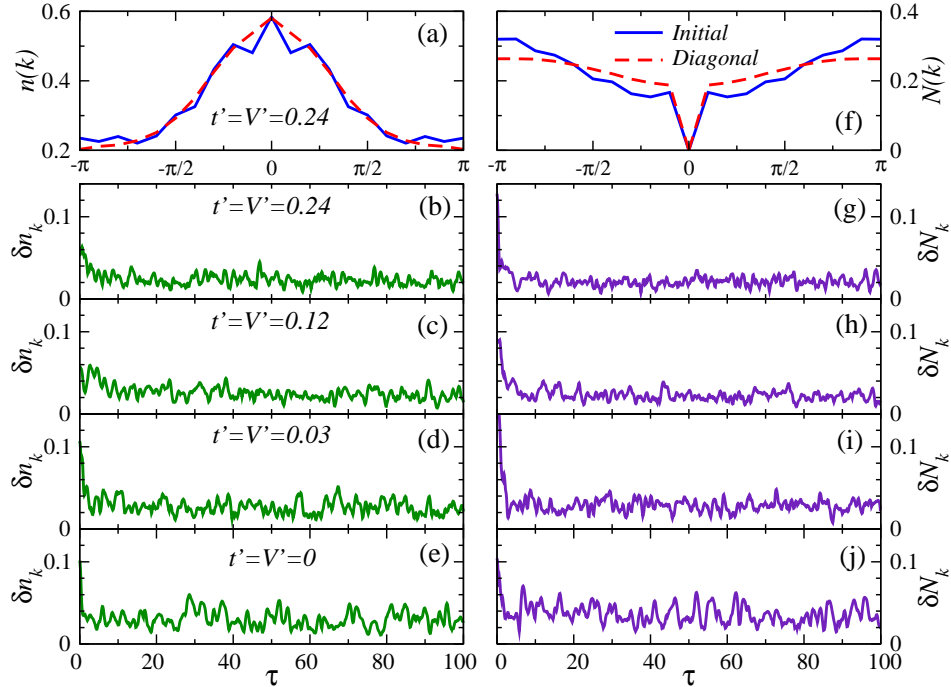}
\end{center}
\vspace{-0.7cm}
\caption{\label{TimeEvolution_L21T3.0}
Quantum quench from $t_{ini}=0.5$, $V_{ini}=2.0$ to $t_{fin}=1.0$, $V_{fin}=1.0$, 
with $t'_{ini}=t'_{fin}=t'$ and $V'_{ini}=V'_{fin}=V'$ in a system with $N_b=7$ 
and $L=21$. The initial state was chosen within the eigenstates of the initial 
Hamiltonian with total momentum $k=0$ in such a way that after the quench the 
system has an effective temperature $T=3.0$ in all cases (see text), as was done 
for Fig.\ \ref{TimeEvolution_L24T3.0}. (a) Initial and diagonal 
ensemble results for $n(k)$ and $t'=V'=0.24$. (b)--(e) Time evolution of 
$\delta n_k$ for $t'=V'=0.24$, 0.12, 0.03, and 0.0 respectively. 
(f) Initial and diagonal ensemble results for $N(k)$ and $t'=V'=0.24$. 
(g)--(j) Time evolution of $\delta N_k$ for $t'=V'=0.24$, 0.12, 0.03, 
and 0.0 respectively.}
\end{figure}

Finally, the effect of changing the system size can be studied by considering 
a smaller system with exactly the same density and effective temperature as the one 
in Fig.\ \ref{TimeEvolution_L24T3.0}. Results for a system with 7 hardcore bosons 
in 21 lattice sites are depicted in Fig.\ \ref{TimeEvolution_L21T3.0}. They exhibit a 
qualitatively similar behavior to the one seen in Fig.\ \ref{TimeEvolution_L24T3.0}, 
but after relaxation $\delta n_k$ and $\delta N_k$ show larger time fluctuations 
and the values around which they fluctuate are also larger than for the system 
with $N_b=8$ and $L=24$. This supports the expectation that for sufficiently large 
system sizes time fluctuations after relaxation will be arbitrarily small and
relaxation will occur towards the exact predictions of the diagonal ensemble.

\onecolumngrid

\twocolumngrid

\subsection{The Microcanonical Ensemble}

\begin{figure}[h!]
\begin{center}
\includegraphics[width=0.485\textwidth,angle=0]{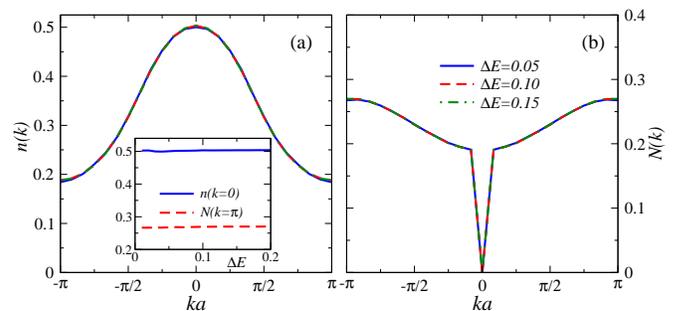}
\end{center}
\vspace{-0.7cm}
\caption{ \label{Micronanonical}
(a) (main panel) Microcanonical momentum distribution function
and (b) density-density structure factor for three different values 
of $\Delta E$. (b) (inset) $n(k=0)$ and $N(k=\pi)$ vs $\Delta E$.
The system has $N_b=8$, $L=24$, and a total energy $E=-3.87$, which corresponds 
to an effective temperature $T=3.0$ for $t'=V'=0.03$ and $t=V=1.0$.}
\end{figure}

The computations of the microcanonical ensemble predictions are done averaging 
over all eigenstates (from all momentum sectors) that lie within a window 
$[E-\Delta E, E+\Delta E]$, where 
$E=\langle \psi_{ini} | \hat{H}_{fin} | \psi_{ini}\rangle$, and we have taken 
$\Delta E=0.1$ in all cases. We have checked that our results are robust in the 
neighborhood of this value of $\Delta E$. In Fig. \ref{Micronanonical}, we show 
(a) the momentum distribution function and (b) the density-density structure factor 
for three different values of $\Delta E$. 
The results for each observable are indistinguishable from each other. Further 
confirmation of this can be obtained looking at the inset in Fig. \ref{Micronanonical}(a), 
where we depict the behavior of $n(k=0)$ and $N(k=\pi)$ vs $\Delta E$ for $0.01<\Delta E<0.2$.
Both quantities can be seen to be almost independent of $\Delta E$ on that window.
Similar results have been obtained for all other values of $t',\,V'$ considered in 
this work.

\subsection{Different Ensembles}

\begin{figure}[!h]
\begin{center}
\includegraphics[width=0.37\textwidth,angle=0]{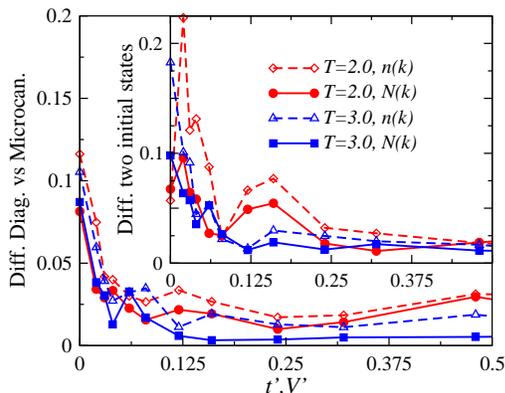}
\end{center}
\vspace{-0.7cm}
\caption{\label{Thermodynamics_L21}
Comparison between the predictions of different statistical ensembles 
for $n(k)$ and $N(k)$ after relaxation. Results are shown for two different 
effective temperatures. (a) (main panel) 
Relative difference between the predictions of the diagonal ensemble 
and the microcanonical ensemble ($L=21$, $N_b=7$).
(a) (inset) Relative difference between the predictions of two diagonal 
ensembles generated by different initial states, selected from the eigenstates 
of a Hamiltonian with $t_{ini}=0.5$, $V_{ini}=2.0$ and a Hamiltonian with 
$t_{ini}=2.0$, $V_{ini}=0.5$. The final Hamiltonian ($t_{fin}=1.0$, $V_{fin}=1.0$) 
and the effective temperature are identical for both initial states.}
\end{figure}

In Fig.\ 2(a) (main text), we compared the predictions of different 
ensembles for our two observables of interest at different effective temperatures, 
but the same system size $N_b=8$ and $L=24$. In order to gain an understanding 
of finite size effects, in Fig.\ \ref{Thermodynamics_L21} we present results for 
identical calculations in smaller systems with $N_b=7$ and $L=21$, which have the 
same density as the ones in Fig.\ 2 (main text). Overall, we find a 
similar qualitative behavior in these smaller systems. However, as expected, 
in the smaller lattices the relative differences are always larger than in 
the larger systems. This is particularly evident away from integrability, where 
in Fig.\ 2(a) (main text) all differences are below one percent while in 
Fig.\ \ref{Thermodynamics_L21} they are two to three times larger than that. 

\subsection{Eigenstate Thermalization Hypotesis}

\begin{figure}[!h]
\begin{center}
\includegraphics[width=0.41\textwidth,angle=0]{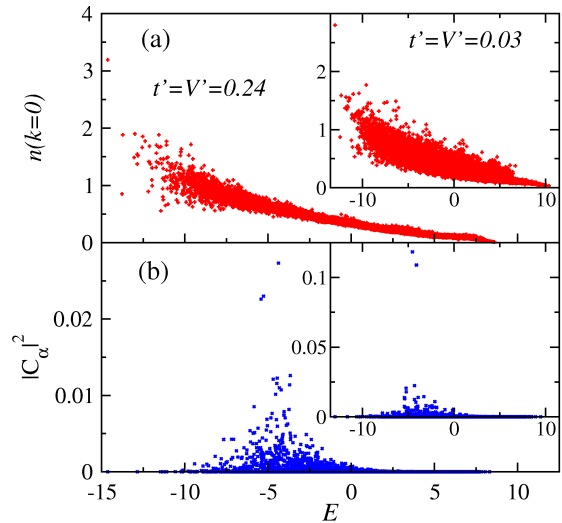}
\end{center}
\vspace{-0.7cm}
\caption{\label{ETH_L21}
(a) $n(k=0)$ as a function of the energy for all the eigenstates of the 
Hamiltonian (from all momentum sectors). (main panel) $t=V=1$ and $t'=V'=0.24$. 
(inset) $t=V=1$ and $t'=V'=0.03$. The system has 21 lattice sites and 7 bosons 
for which the total Hilbert consists of 116,280 states.
(b) Distribution of $|C_{\alpha}|^{2}$ for the quench from 
$t_{ini}=0.5$, $V_{ini}=2.0$ to $t_{fin}=1.0$, $V_{fin}=1.0$ for an
effective temperature $T=3.0$.
(main panel) $t=V=1$ and $t'=V'=0.24$, where 
$E=\langle\psi_{ini}\vert \widehat{H}_{fin}\vert \psi_{ini}\rangle=-3.85$. 
(inset) $t=V=1$ and $t'=V'=0.03$, where 
$E=\langle\psi_{ini}\vert \widehat{H}_{fin}\vert \psi_{ini}\rangle=-3.44$.}
\end{figure}

In Fig.\ \ref{ETH_L21}(a), we show the eigenstate expectation values of $n(k=0)$ 
as a function of the energy of the eigenstate for a system with $L=21$ and $N_b=7$. 
The values of $t,\,V,\,t'$ and $V'$ in the main panel and the inset correspond to 
the exact same values presented in Fig.\ 4(a) (main text) for $L=24$ and $N_b=8$. 
The behavior of $n(k=0)$ is 
very similar in both systems in which the lattice size and the total filling are the 
only difference, as we have kept the density constant. These results exemplify the 
robustness of ETH or its failure for our finite systems. Notice that in Fig.\ 4 (main text)
and Fig.\ \ref{ETH_L21}, we are showing results for the eigenstate expectation values 
of $n(k=0)$ in the full spectrum (including all momentum sectors). The validity of 
ETH in the full expectrum explains why our microcanonical calculations, in which we 
included the contributions from all momentum sectors, provide a correct prediction 
for the outcome of the diagonal ensemble even though in the latter ensemble only 
states with zero total momentum have a finite weight. 

In Fig.\ \ref{ETH_L21}(b), we show the distribution of $|C_\alpha|^2$ for the same 
Hamiltonian parameters as in Fig.\ 4(b) (main text), but once again for a smaller system with 
$L=21$ and $N_b=7$, and also for a higher effective temperature of the time evolving 
state. In Fig.\ 4(b) (main text), we showed results for $T=2.0$ and in Fig.\ \ref{ETH_L21}(b),
we are showing results for $T=3.0$. Qualitatively, one can see that there is not much difference
between those two cases. The distribution of $|C_\alpha|^2$ is in stark contrast with the 
weight given to each eigenstate expectation value in the microcanonical and canonical 
distributions. $|C_\alpha|^2$ vs $E_\alpha$ is a distribution that strongly depends on 
the initial conditions of the system, and $|C_\alpha|^2$ exhibits strong fluctuations 
between contiguous eigenstates. The only generic feature one can see by comparing 
Fig.\ 4(b) (main text) with Fig.\ \ref{ETH_L21}(b) is that in both cases the distribution 
of $|C_\alpha|^2$ is peaked around the mean energy of the time evolving state.

\subsection{Deviations from ETH}

To have a quantitative understanding of how ETH breaks down as one approaches integrability,
we computed the average relative deviation of the eigenstate expectation values with 
respect to the microcanonical prediction, $\Delta^{mic}n(k=0)$ and $\Delta^{mic}N(k=\pi)$. 
For any given energy of the microcanonical ensemble, these quantities are computed as
\begin{equation}
 \Delta^{mic}n(k=0)=\dfrac{\sum_{\alpha}|n_{\alpha \alpha}(k=0)-n_{mic}(k=0)|}{N_{states}\,n_{mic}(k=0)}
\end{equation}
and
\begin{equation}
\Delta^{mic}N(k=\pi)=\dfrac{\sum_{\alpha}|N_{\alpha \alpha}(k=\pi)-N_{mic}(k=\pi)|}{N_{states}\,N_{mic}(k=\pi)}
\end{equation}
where $n_{\alpha \alpha}(k=0)$ and $N_{\alpha \alpha}(k=\pi)$ are the eigenstate expectation values of 
$n(k=0)$ and $N(k=\pi)$, respectively, and $n_{mic}(k=0)$ and $N_{mic}(k=\pi)$ are the microcanonical 
expectation values at any given energy $E$. The sum over $\alpha$ contains all states 
with energies in the window $[E-\Delta E, E+\Delta E]$, and $N_{states}$ is the number of 
states in the sum. As discussed before $\Delta E=0.1$. 

\begin{figure}[!h]
\begin{center}
\includegraphics[width=0.485\textwidth,angle=0]{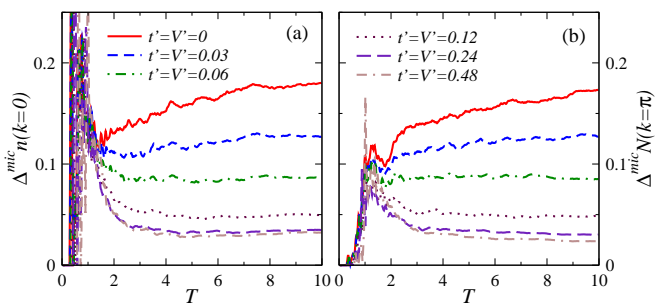}
\end{center}
\vspace{-0.7cm}
\caption{\label{DeviationfromETH_L21}
Average relative deviation of eigenstate expectation values with 
respect to the microcanonical prediction as a function of the 
effective temperature of the microcanonical ensemble. Results are presented 
for: (a) $n(k=0)$ and (b) $N(k=\pi)$. These systems 
have 21 lattice sites, 7 bosons, and $t=V=1.0$. An effective 
temperature $T=2$ in these plots corresponds to an energy 
$E_{mic}=-5.25$ for $t'=V'=0.03$ and 
$E_{mic}=-6.05$ for $t'=V'=0.24$ and, 
$T=10$ corresponds to an energy 
$E_{mic}=-0.75$ for $t'=V'=0.03$ and 
$E_{mic}=-0.76$ for $t'=V'=0.24$
(see the corresponding parts of the spectrum in Fig.\ \ref{ETH_L21}).}
\end{figure}

In order to compare results for different Hamiltonian parameters, we have plotted 
$\Delta^{mic}n(k=0)$ and $\Delta^{mic}N(k=\pi)$ as a function of the effective temperature $T$,
corresponding to an energy $E$ of the microcanonical ensemble. Having the energy of the microcanonical 
ensemble $E$, $T$ can be computed from $E=Z^{-1}\textrm{Tr}\left\lbrace {\hat{H} \exp(-\hat{H}/{k_B T})}\right\rbrace$,
where $Z=\textrm{Tr}\left\lbrace {\exp(-\hat{H}/{k_B T})}\right\rbrace$. 
Notice that here this effective temperature is only used as an auxiliary tool for assessing how 
far from the ground state these systems are, and to have a unique energy scale window for 
comparing different observables independently of the Hamiltonian parameters $t'$ and $V'$, 
which change the ground state energy and level spacing.

Results for $\Delta^{mic}n(k=0)$ and $\Delta^{mic}N(k=\pi)$ are shown in 
Fig.\ \ref{DeviationfromETH_L21}(a) and \ref{DeviationfromETH_L21}(b), respectively, for a system
with $L=21$ and $N_b=7$. The results depicted in Fig.\ \ref{DeviationfromETH_L21} are 
qualitatively similar to those in Fig.\ 5 (main text). 
Below $T\simeq 1.5$ fluctuations are in general large and nonmonotonic. For $T\gtrsim 1.5$, one can 
see that $\Delta^{mic}$ for $n(k=0)$ and $N(k=\pi)$ continuously increase, at any given temperature,
as one approaches the integrable point, i.e., there is no abrupt transition as $t',V'\rightarrow 0$.

\begin{figure}[!h]
\begin{center}
\includegraphics[width=0.485\textwidth,angle=0]{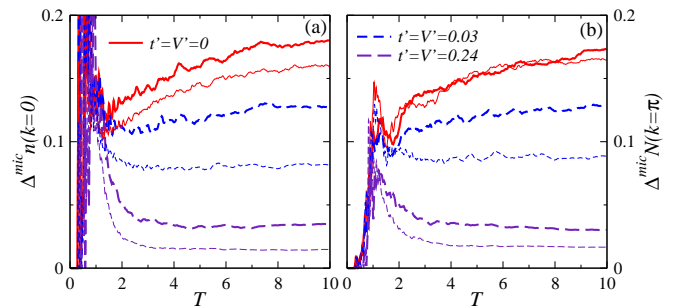}
\end{center}
\vspace{-0.7cm}
\caption{\label{DeviationfromETH_comparison}
Comparison between the average relative deviation of eigenstate expectation values 
with respect to the microcanonical prediction for two different system sizes. 
$L=21$ and $N_b=7$ (thick lines), and $L=24$ and $N_b=8$ (thin lines). 
Results are presented for: (a) $n(k=0)$ and (b) $N(k=\pi)$. 
In all cases $t=V=1.0$.}
\end{figure}

A direct comparison between some of the results presented in Fig.\ 5 (main text)
and Fig.\ \ref{DeviationfromETH_L21} is depicted in Fig.\ \ref{DeviationfromETH_comparison}.
There, one can see that $\Delta^{mic}$ in general reduces as one increases 
the system size [except for $\Delta^{mic}N(k=\pi)$ when $t'=V'=0$]. The relative 
change is more pronounced away from integrability. These results do not allow us to discriminate
between two possible scenarios for the breakdown of thermalization in quantum systems. A first 
scenario in which ETH may hold arbitrarily close to the integrable point as the system size increases, 
and another possible scenario in which in the thermodynamic limit there may be a critical value of 
$t',\,V'$ (the integrability breaking terms) below which ETH does not hold and above which it does.
This question will need to be addressed in future works.

\begin{thebibliography}{99}
\bibitem{temperature}
Given an energy $E$, the effective temperature $T$ is defined by the
expression $E=Z^{-1}\textrm{Tr}\left\lbrace {\hat{H} \exp(-\hat{H}/{k_B T})}\right\rbrace$, 
where $\hat{H}$ is the Hamiltonian for which $E$ is measured, 
$Z=\textrm{Tr}\left\lbrace {\exp(-\hat{H}/{k_B T})}\right\rbrace$, and $k_B$
is the Boltzmann constant. The trace runs over the full spectrum.
\bibitem{smooth} Its relation with the crossover between Poissonian 
and Wigner-Dyson level statistics is discussed in Ref.\ \cite{santos09}.
\end{thebibliography}
\end{document}